# Archean Methane Cycling and Life's Co-Evolution: Intertwining Early Biogeochemical Processes and Ancient Microbial Metabolism

Saleheh Ebadirad[1], Timothy W. Lyons[1], Gregory P. Fournier[2]

[1] Department of Earth and Planetary Sciences, University of California, Riverside, CA, United States

[2] Department of Earth, Atmospheric and Planetary Sciences, Massachusetts Institute of Technology, Cambridge, MA, United States

## Abstract:

This chapter explores the key carbon compounds that shaped the Archean biogeochemical cycle, delineating their substantial impact on Earth's primordial atmospheric and biospheric evolution. At the heart of the Archean carbon cycle were carbon dioxide and methane, which served as key regulators of Earth's early climate. Particular emphasis is placed on methane cycling, encompassing both abiotic methane production and consumption, as well as their biotic counterparts—methanogenesis and methanotrophy. These ancient microbial pathways not only shaped methane fluxes but were also tightly interwoven with Earth's evolving redox state. We provide a comprehensive exploration of the intertwined evolution of Earth's geochemical environment and microbial life. The interdisciplinary approach of this chapter not only sheds light on the complex dynamics of Earth's early methane cycling but also offers critical insights that could inform the search for life beyond our planet, thereby marking a contribution to Earth sciences, astrobiology, and related fields.

## Introduction:

The Archean carbon cycle was shaped by a complex interplay of processes, integrating both geological and biological components. Geological processes such as weathering, carbonate chemistry, and deep-sea circulation, along with the emergence and evolution of diverse metabolic activities in early microbes and their ecological interactions, were key players. Nutrient cycles, acting as a bridge between these geological and biological processes, alongside inherent feedback mechanisms, further contributed to shaping Earth's ancient environment (Homann, 2019; Lepot, 2020; Lyons et al., 2015; Lyons, Tino et al., 2024). At the core of this cycle were the primary forms of inorganic single-carbon molecules, carbon dioxide ($CO_2$), the most oxidized form of carbon, and methane ($CH_4$), the most reduced form. Carbon dioxide is a fundamental component in a myriad of abiotic reactions on Earth and serves as an essential substrate for primary producers, making it a cornerstone molecule in the carbon cycle. Its role extends from being a greenhouse gas that regulates Earth's climate to being a critical carbon source for chemotrophs and photosynthetic organisms, which convert $CO_2$ into organic matter, thus fueling the terrestrial and marine food webs. As such, it plays a multifaceted role in Earth's carbon cycle, with its sources and sinks tied to a blend of abiotic and biotic processes (Catling & Zahnle, 2020; Lepot, 2020). During the

Archean Eon spanning from approximately 4 to 2.5 billion years ago (Ga), Earth experienced significant geological and biological changes. Within this timeframe, $CH_4$ is also believed to have played a more important role in shaping the Earth's atmosphere and its environments than it does in the present oxidizing world, including first-order controls on climate. Along with $CO_2$, it likely contributed to maintaining surface temperatures within a range conducive to supporting liquid water, thereby facilitating the emergence and persistence of a biosphere (Catling & Zahnle, 2020; Charnay et al., 2020). This chapter delves into the pivotal role of $CH_4$ cycling and related processes throughout the Archean.

We describe the potential sources and sinks of $CH_4$ during the Archean, focusing on the geological and biological mechanisms that contributed to its production and consumption. Various lines of evidence, including isotopic signatures like $\delta^{13}C$ values in Archean rocks and genomic data, suggest the presence of $CH_4$ and methane-cycling organisms very early in Earth history (Bell et al., 2015; Lepot, 2020; Wolfe & Fournier, 2018). In this chapter, special attention will be given to methanogenesis, the biological production of $CH_4$ by archaea, and methanotrophy, the microbial consumption of $CH_4$. These processes are central to the $CH_4$ cycle and likely had profound impacts on the broader carbon cycle and Earth's early climate system. By exploring geological and biological factors, we aim to elucidate how geochemical records from ancient rocks and genomic evidence from microbial communities can help us reconstruct early Earth conditions, thereby deepening our understanding of paleoclimatology and paleoceanography. The interdisciplinary nature of this investigation, which bridges geological and biological sciences, is crucial for a comprehensive grasp of Archean $CH_4$ cycling. Furthermore, the insights gained not only expand our knowledge of Earth history but also extend beyond our planet, offering valuable perspectives for identifying and interpreting habitable environments elsewhere. This approach should enable scientists to target exoplanets with potential for life more effectively, guiding future missions in the search for extraterrestrial biosignatures.

## Overview of Earth's early atmosphere and environmental conditions:

For more than 4 billion years, Earth has maintained a habitable environment that included liquid water oceans, as suggested by isotopic signatures found in zircon crystals dating back to the Hadean (Cameron et al., 2024; Wilde et al., 2001). The subsequent Archean eon, which lasted from 4 to 2.5 Ga, accounts for about one-third of our planet's history. Numerous studies and paleoclimate models suggest that the compositions of Earth's atmosphere and oceans during this period were fundamentally different from those of today (Albarede et al., 2020; Catling & Zahnle, 2020; Lyons et al., 2014). The Archean atmosphere was primarily composed of carbon dioxide ($CO_2$) and nitrogen ($N_2$), but it also contained significant quantities of reducing gases such as carbon monoxide (CO), hydrogen ($H_2$), and methane ($CH_4$), and was devoid of oxygen gas ($O_2$) (Catling & Zahnle, 2020; Kasting, 2010).

In addition to these compositional differences, the magnitude and distribution of solar insolation significantly influences Earth's climate. Early in Earth's history, insolation levels were lower but have progressively increased over time. Average surface temperatures during the Archean are estimated to have ranged between 0° and 40°C, which aligns with evidence of occasional ice ages (Catling & Zahnle, 2020). However, some studies suggest that surface and ocean temperatures during the Archean could have reached up to 55° to 85°C (Blake et al., 2010; Knauth, 2005), which is significantly warmer than today's conditions. Moreover, pH levels in these ancient waters are estimated to have been roughly $6.6^{+0.6}_{-0.4}$ at 4.0 Ga and $7.0^{+0.7}_{-0.5}$ at 2.5 Ga (compared to ca. 8.0 today), i.e. moderately acidic to relatively neutral (Krissansen-Totton et al., 2018a). These conditions were likely conducive to diverse microbial life forms that were adapted to moderately high temperatures and varying—though roughly circumneutral—pH levels during the Archean.

Moreover, the Archean oceans, under anoxic and reducing conditions, had a distinct composition of metals reflecting the overall low oxygen levels and more mafic crustal compositions compared to today's. This included low levels of copper (Cu), zinc (Zn), and molybdenum (Mo) but enrichments in iron (Fe) and other metals such as nickel (Ni), manganese (Mn), and cobalt (Co) (Anbar, 2008; Saito et al., 2003). These metals are vital for biogeochemical processes, acting as key nutrients and cofactors in many microbial metabolisms. Studying their specific abundances in ancient rocks provides valuable insights into ancient redox conditions and potential productivity levels. These insights are crucial for reconstructing paleoenvironmental conditions and the magnitudes and patterns of activity in the early biosphere (Dupont et al., 2010; Scott et al., 2008; Tribovillard et al., 2006).

Initially, Earth's carbon cycle was governed by abiotic processes; such as volcanic outgassing; carbon fluxes from seeps; and the cycling of carbon through various mechanisms, including subduction zone processes, weathering, and carbonate formation (Hayes & Waldbauer, 2006; Sleep & Zahnle, 2001). Over time, this cycle has undergone complex evolution with the advent of life and subsequent evolution of organisms specifically capable of chemosynthesis, photosynthesis, carbon fixation, and skeletonization, introducing biotic contributions to carbon cycling (Lyons et al., 2015; Lyons et al., 2021). Geological evidence suggests the presence of microbial communities as early as approximately 3.5 to 3.8 Ga, and potentially even earlier (e.g., Bell et al., 2015; Javaux, 2019). The early diversity of microbial life, including methanogens and methanotrophs, played a crucial role in shaping the planet's habitability and influencing Earth's biogeochemical cycles. They contributed significantly to the carbon and methane cycling on the early Earth, fundamentally impacting the development of the planet's atmosphere and its ability to support life (Lepot, 2020; Lyons et al., 2015; Lyons, Tino et al., 2024). The subsequent evolution of oxygenic photosynthesis and aerobic metabolisms in the later Archean marked a turning point in Earth history. This significant evolutionary shift led to the Great Oxygenation Event (GOE), which occurred roughly between 2.4 and 2.3 Ga (e.g., Fournier et al., 2021; reviewed in Lyons et al., 2014; Lyons, Tino et al., 2024). The GOE represents a transition to greater redox potential, leading to a new regime of biogeochemical cycling. This shift significantly reduced the role of $CH_4$ in climate and carbon cycling, altering the states and dynamics of Earth's atmospheric and oceanic systems and paving the way for the emergence of more complex life forms.

## Faint Young Sun Paradox

Astronomical studies have estimated that solar luminosity during the early Precambrian was approximately 20-30% lower than current levels, attributed to a higher proportion of hydrogen relative to helium in the Sun's core at that time (e.g., Gough, 1981; Kasting & Ono, 2006; Kasting, 2010). If so, Earth should have been entirely frozen during the first half of its existence; however, it was not. Geological records suggest that despite the faint young Sun, Earth was warm enough to maintain a liquid hydrosphere and support life throughout most of its early history, experiencing only occasional glaciations (Charnay et al., 2020; Feulner, 2012). These records include, but are not limited to, evidence of liquid oceans dating back to around 4.4 to 4.3 Ga, as preserved in the oxygen isotope ratios of detrital zircons from Jack Hills, Western Australia (Mojzsis et al., 2001; Valley et al., 2014; Wilde et al., 2001). There are also potential signs of early life, indicated by $\delta^{13}C$ ratios consistent with biogenic carbon detected in 4.1 Ga graphite within a detrital zircon from the Archean Jack Hills sandstone, which has been attributed to an unknown microbial source (Bell et al., 2015). This interpretation would indicate the presence of autotrophic metabolic pathways comparable to those of modern organisms.

Such hypotheses can potentially be tested by examining the evolutionary history of genes and organisms preserved within genomes by inferring their phylogenetic relationships. Genome sequences and phylogenies also allow us to estimate divergence times using relaxed molecular clock models, fossil calibrations, and other dating information. This approach, together with geochemical data, allows us to reconstruct a timescale for microbial life and its metabolic processes (Fournier et al., 2021; Lyons, Tino et al., 2024). Consequently, a growing body of phylogenetic studies and well-calibrated molecular clock analyses further bolster the hypothesis that ancient life forms, such as chemotrophs (including methane-cycling microbes) and phototrophs (including ancestral cyanobacterial lineages), emerged very early in Earth history, during the Archean Eon or perhaps even predating it (Battistuzzi, et al., 2004; Fournier et al., 2022; Szöllõsi et al., 2022).

These collective findings present a challenge to our understanding of early Earth's habitability. Despite the Sun's fainter luminosity during the Archean Eon, these discoveries suggest that Earth managed to sustain conditions conducive to life, including the persistence of liquid water and suitable temperatures. While the prevailing interpretation of the Archean geological record suggests the planet was not glaciated on a global scale, difficulties in accurately determining the paleolatitudes of Archean sedimentary basins raise questions about the assumption of uniformly high global temperatures (Spalding & Fischer, 2019). The likelihood of enduring liquid oceans beneath a faint Sun has prompted investigations into the mechanisms that facilitated Earth's ability to sustain habitable conditions during the Archean, which could offer valuable insights into the early evolution of our planet and the potential for Archean Earth-like habitable planets beyond our solar system (Olson et al., 2018). Several hypotheses have been proposed to explain how early Earth remained warm despite the Faint Young Sun.

One such hypothesis is that the energy released during Earth's accretion and subsequent radioactive decay of planetary materials could have served as additional heat reservoirs. Another posits that strong tidal heating, resulting from a closer Moon, could have facilitated the presence of liquid water (Heller et al., 2021). A more recent hypothesis proposes that hypervelocity dust, driven by coronal mass ejections from super solar flares, may have entered Earth's atmosphere and contributed significantly to its warming (Rodriguez et al., 2024). Additionally, geothermal heat fluxes, hydrothermal activity, and volcanism may have played a crucial role in providing the primordial heat necessary for sustaining Earth's early habitability (Ojha et al., 2020). Other hypotheses consider ways in which the early Earth would maintain warmer temperatures by having a lower albedo. For example, hypothesized higher ocean salinity during the Archean would have depressed the freezing point of seawater, inhibiting sea ice formation. Reduced sea ice extent lowers the albedo, amplifying warming through the positive ice-albedo feedback mechanism (Fofonoff & Millard, 1983; Olson et al., 2022).

However, the greenhouse effect is likely to be the most important factor in understanding the surface temperature of the early Earth given the Faint Young Sun. Many greenhouse gases would have been present in Earth's primordial atmosphere, including $CO_2$, $CH_4$, $H_2O$ and possibly minor $NH_3$. Carbon dioxide is presumably the most important of these greenhouse gases, although it is not known if $CO_2$ concentrations at the time were sufficient in themselves to counteract the Faint Young Sun. Higher concentrations of other greenhouse gases, such as ammonia ($NH_3$) and $CH_4$, may also have been necessary (Pavlov et al., 2000; Sagan & Mullen, 1972). In an atmosphere without $O_2$, reduced greenhouse gases like $CH_4$ and $NH_3$ might have been more abundant and could have persisted for longer periods (Kasting & Catling, 2003). However, estimated abiotic sources of $NH_3$, or possible biotic sources following the emergence of life, were likely insufficient to generate significant greenhouse warming during the early Archean (Kasting, 1982). Furthermore, $NH_3$ is photochemically unstable and undergoes irreversible conversion to $N_2$ and $H_2$ rapidly, and thus could not be sustained against UV photolysis in the absence of $O_2$ and therefore UV-shielding ozone, which would have limited its long-term presence in the early atmosphere (Kasting, 1982; Kuhn & Atreya, 1979). But what about $CH_4$? Could this potent greenhouse gas have been a climate savior in the Archean?

## A Methane-Rich Archean Atmosphere?

Arguably, $CH_4$ was a key atmospheric component throughout a significant portion of Earth's Precambrian climate system. $CH_4$ could have been generated through various chemical reactions in geological settings. Some of this reduced gas likely originated abiotically from high-temperature magmatic activities in geothermal and volcanic regions, or through relatively low-temperature reactions between gas, water, and rock in continental settings or at shallow ocean depths on early Earth (Etiope & Lollar, 2013). During the early stages of Earth's history, the prevalence of ultramafic oceanic crust, primarily composed of komatiite, was conducive to serpentinization reactions. These reactions, through the oxidation of iron in rocks, could produce $H_2$, which in turn could fuel methane production and increase atmospheric $CH_4$ concentrations (Tamblyn & Hermann, 2023). Additionally, $CH_4$ production in the primordial environment could have occurred on the surface of metallic catalysts supplied by meteorite impacts (Sekine et al., 2003). Earlier in Earth history, before the Archean, atmospheric $CH_4$ could have favored the formation of prebiotic molecules, although the overall $CH_4$ levels were likely low and transient (Zahnle et al., 2020).

As Earth's biosphere evolved during the interval between the emergence of life and the GOE, the anoxic atmosphere and oceans were likely dominated and heavily influenced by anaerobic microbes. Perhaps the most significant of these were methanogens, which could have had a substantial impact on the Archean atmosphere by producing high levels of $CH_4$ (Catling & Zahnle, 2020; Laakso & Schrag, 2019; Lepot, 2020; Olson et al., 2016). Methanogenic sources of $CH_4$ were likely a major contributor to the greenhouse effect at this time, and are estimated to have been responsible for approximately 10-12°C of surface warming (Kasting, 2004; Kasting, 2010). Apart from $CH_4$, higher concentrations of $CO_2$ and other greenhouse gases such as water vapor and possibly nitrous oxide ($N_2O$) may have also played an important role during the Archean. Conversely, decreasing concentrations of $CH_4$ and other greenhouse gases, combined with other

factors—particularly early atmospheric oxygenation—may have led to lower surface temperatures and contributed to the severe, prolonged ice ages during the Paleoproterozoic (Charnay et al., 2020; Haqq-Misra et al., 2008; Poulton et al., 2021; Airapetian et al, 2018). A decrease in the amount of atmospheric $CH_4$ produced by methanogens may have coincided with the gradual rise of atmospheric $O_2$. One cause for this correlation may have been oxygen toxicity, restricting the niche of obligate anaerobe methanogens to remaining anoxic habitats, such as sediments (Catling et al., 2007; Konhauser et al., 2009).

However, an overabundance of atmospheric $CH_4$ may have had other consequences for the early Earth. Several studies have highlighted that before the GOE, in the absence of ozone shielding, higher levels of solar UV irradiation and $CH_4$ might have triggered the formation of a high-altitude organic haze primarily composed of hydrocarbon particles such as ethane ($C_2H_6$) (Arney et al., 2016; Haqq-Misra et al., 2008). The density or optical thickness of this organic haze in the Archean atmosphere is directly influenced by the atmospheric $CH_4$-to-$CO_2$ ratio. When this ratio in the atmosphere exceeds approximately 0.1, the organic haze becomes denser and more optically thick (Haqq-Misra et al., 2008). One intriguing speculation for the Archean Earth is its characterization as the "pale orange dot," a term reflecting this widely held belief that a methane-rich, hazy atmosphere may have imparted a reddish-orange hue to the planet, akin to the appearance of Titan today, rather than the pale blue of modern Earth. This distinct coloration would be caused by the absorption and scattering effects of atmospheric haze particles, particularly in the visible and near-infrared spectra (Arney et al., 2016; Zahnle et al., 2020).

Such an organic haze would have presented both opportunities and challenges for early Earth's environment. For instance, it would have increased planetary habitability by reducing surface UV flux by about 97%. This UV shielding might have enabled terrestrial microbial life to thrive during the later stages of Earth history in the Archean (Arney et al., 2016; Trainer, 2013). As a potential obstacle to habitability, such an organic haze could have induced surface cooling through a pronounced 'anti-greenhouse' effect (Arney et al., 2016). This cooling occurs because haze particles scatter and reflect a significant portion of incoming solar radiation, preventing it from reaching the Earth's surface. However, organic hazes are thought to be self-limiting. At elevated $CH_4$ ratios relative to $CO_2$, haze thickness and corresponding surface temperatures reach a state of equilibrium. This stabilization occurs as UV self-shielding reduces $CH_4$ photolysis, thus inhibiting further haze formation. Moreover, oxygen radicals produced as a result of $CO_2$ photolysis can inhibit haze formation, thereby preventing catastrophic cooling of the planet. This self-shielding process helps maintain a steady state that can permit sufficient planetary warming to maintain habitability. Consequently, a more balanced greenhouse gas inventory could provide a better solution to the Faint Young Sun paradox (Arney et al., 2016; Wolf & Toon, 2013).

## Sources and Sinks of Methane During the Archean

Methane plays a crucial role at the intersection of geological and biological processes, bridging the abiotic and biotic worlds (Etiope & Lollar, 2013; Garcia et al., 2022). As a potent greenhouse gas, $CH_4$ acts as a significant atmospheric component influencing Earth's climate, and as discussed, atmospheric $CH_4$ could have contributed to maintaining sufficient warmth on the Archean Earth prior to the GOE (Haqq-Misra et al., 2008; Kasting et al., 1983). However, the fact that photochemical reactions could have destroyed atmospheric $CH_4$ in the absence of ozone shielding suggests that if significant levels of CH$_4$ persisted in the atmosphere during the Archean, it must have been continuously produced and replenished. It has been proposed that biotic processes would likely have been necessary to achieve and sustain the levels of CH$_4$ required to contribute to a methane greenhouse, although this hypothesis remains largely untested ([Arney et al., 2016](#); [Zahnle et al., 2010](#)). Such biotic sources would only exist after the origin and evolution of microbial life and so cannot explain primordial CH$_4$ greenhouse contributions, but nevertheless may have been important for longer-term habitability. Methane also has biotic sinks that play a primary role in the modern carbon cycle, and a similar scenario likely existed during the Archean. In today's environment, CH$_4$ is oxidized by both aerobic and anaerobic methanotrophs. In anoxic sediments, archaeal CH$_4$ consumption can occur through reverse methanogenesis, using external electron acceptors such as sulfate, nitrate/nitrite, and various metal oxides. In oxic environments, aerobic bacterial methanotrophs consume CH$_4$ via using O$_2$ as a terminal electron acceptor through a different mechanism ([Hallam et al., 2004](#); [Timmers et al., 2017](#)). This process is distinct from anaerobic methane oxidation by ANaerobic MEthanotrophs (ANME) and relies on different genes and pathways ([Borrel et al., 2019](#); [Garcia et al., 2022](#); [Kalyuzhnaya, 2019](#); [Smith et al., 2011](#))

Investigating the isotopic composition of CH$_4$ is an important means of distinguishing abiotic and biotic sources ([Etiope & Lollar, 2013](#)). Clumped isotope analysis, measuring the multiply substituted isotopologues of methane, reveals the temperature at which CH$_4$ forms. This temperature-dependent fractionation provides a means to differentiate between biotic and abiotic processes and serves as a geothermometer. Currently, the known upper temperature limit for hyperthermophilic methanogenic life is approximately 122°C ([Takai et al., 2008](#)), whereas thermogenic (abiotic) CH$_4$ forms predominantly at temperatures higher than ~150°C. Thus, clumped isotope analysis can provide independent and informative constraints on CH$_4$ formation temperatures, which can help distinguish between biogenic and thermogenic sources ([Giunta et al., 2019](#); [Stolper et al., 2014](#)). Figure 1 provides a comprehensive depiction of the sources and sinks of CH$_4$ during Archean.

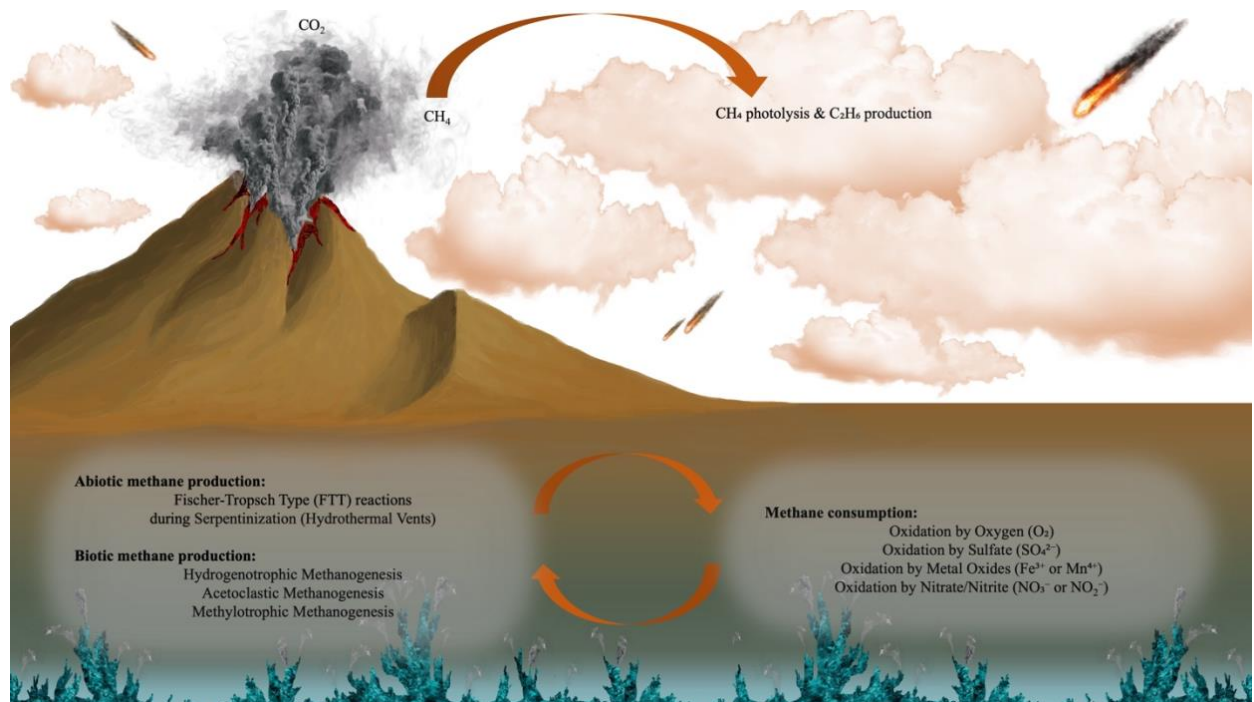

**Fig. 1: Methane Cycle During the Archean.** High-temperature magmatic processes and low-temperature gas-water-rock reactions, such as Fischer-Tropsch Type (FTT) reactions during serpentinization near hydrothermal vents, were probably the most important Archean abiotic sources of $CH_4$. Additionally, some amount of primordial $CH_4$ could have been released as a result of meteorite impacts. As time passed, different microbial methanogenesis pathways, including hydrogenotrophic, acetoclastic, and methylotrophic, started to contribute to Archean $CH_4$ content. Through photochemical reactions, $CH_4$ was converted to more complex hydrocarbons, creating a hazy atmosphere. Abiotic and biotic oxidation processes were also likely present on Precambrian Earth to consume $CH_4$. Oxidation of $CH_4$ could have happened through processes such as anaerobic sulfate-dependent, metal-dependent, or nitrate/nitrite-dependent anaerobic methanotrophy pathways. Aerobic methanotrophy also contributed to oxidation, likely following the evolution of oxygenic photosynthesis and the availability of $O_2$ as an electron acceptor.

- **The Role of Geological Processes**

Prebiotic chemistry on the early Earth may have required the presence of reduced gases such as $H_2$, $CH_4$, and $NH_3$. However, geological evidence suggests that Earth's mantle has consistently been relatively oxidized, primarily emitting gases like $CO_2$, $N_2$, and $H_2O$ (Zahnle et al., 2020). This oxidized mantle is thought to have resulted from the formation of the core, which likely occurred within the first 30-200 million years following Earth formation. The hafnium-tungsten ($^{182}Hf$-$^{182}W$) decay system serves as a crucial chronometer for dating this process. The segregation of a metal core from silicates could have led to an oxidized mantle soon after core formation (Carlson et al., 2015; Nimmo & Kleine, 2015). The evolution of the mantle's redox state is closely tied to the oxidation state of the early atmosphere (Kasting et al., 1993); however, the precise oxidation state of the early mantle remains a topic of debate (Kasting, 2014; Yang et al., 2014). This key parameter can regulate processes such as planetary differentiation, mantle melting, the speciation of C-O-H-bearing fluids, and volcanic outgassing. Thus, a shift from a low to a high oxygen state in the mantle could have significantly impacted the cycling of volatile substances, such as $CH_4$-rich fluids. If oxygen levels in the upper mantle increased during the very early stages of Earth's history, water vapor and $CO_2$ would have become the predominant volcanic gases (Frost & McCammon, 2008; Nicklas et al., 2018), while $CH_4$ emissions likely decreased significantly.

One possible solution to this apparent contradiction may be provided by the reducing power of material that accreted onto Earth through large impacts following the Moon-forming impact (Zahnle et al., 2020). However, if life originated at hydrothermal vents, as some theories suggest, the composition of atmosphere might have been less critical, given that vent environments can provide necessary reductants independent of the global atmospheric conditions.

Modeling studies support that a pre-impact atmosphere would have contained high levels of $CO_2$ and some $N_2$. This $CO_2$-dominated atmosphere is thought to have prevailed during the Hadean and was associated with a liquid ocean (Zahnle et al., 2020). During the early stages of Earth's evolution, intense bombardment by small celestial bodies, with both deleterious and favorable effects, profoundly impacted the young planet's environment and the development of early life (Bottke & Norman, 2017). These impacts could have introduced catalytically active iron and nickel, which were globally ejected and likely re-entered the atmosphere. Friction with atmospheric gases could have heated these metals and potentially facilitated Fischer-Tropsch catalysis—a process for converting CO and $H_2$ into hydrocarbons, primarily $CH_4$. This primordial $CH_4$, upon photodissociation in the upper atmosphere, could lead to the formation of hydrogen cyanide (HCN), a crucial compound for synthesizing amino acids and purines. Thus, large meteor impacts could have played a significant role in early Earth's prebiotic chemistry by forming a transient atmosphere rich in reduced gases, including $CH_4$. Such an impact-driven reduced atmosphere could have contributed to the formation of life's building blocks (Sekine et al. 2003; Wogan et al, 2023; Zahnle et al., 2020). Under certain conditions, such as those stemming from frequent impacts, $CH_4$ could have persisted for extended periods in an atmosphere. However, $CH_4$ is prone to oxidation or polymerization into more complex hydrocarbons, and photochemical processes would have depleted $CH_4$ levels (Kasting et al., 1983; Zahnle, 1986). Thus, even in the most optimistic scenario, these impacts alone on a planet devoid of biological production would have been insufficient to sustain the $CH_4$ levels predicted in later stages of Earth history. Nevertheless, abiotic sources of $CH_4$ would have contributed to maintaining temperatures above freezing before the advent of methanogenic archaea (Kress & McKay, 2004).

After the Moon-forming impact, Earth possessed a core of similar size to today's, a silicate mantle, and temperate surface conditions. The subsequent emergence of plate tectonics introduced additional geological processes that could have modulated $H_2$ fluxes and the production of $CH_4$. The precise emergence time and the mechanisms of plate tectonics and its subsequent evolution remain fundamental yet challenging questions in geology (Brown et al., 2020; Tutolo et al., 2020). Various studies have proposed that plate tectonics commenced at different times, ranging from the early Hadean to approximately 700 million years ago (Myr). Regardless, there is evidence that Archean cratons stabilized at varying times in different regions, between 3.1 and 2.5 Ga. During this timeframe, the composition of juvenile continental crust transitioned from mafic compositions to more intermediate ones (Hawkesworth et al., 2020; Lee et al., 2016). Mineral inclusions in diamonds provide 3.5 billion years of mantle data, revealing a compositional shift around 3.0 Ga. Before 3.2 Ga, only peridotitic diamonds formed, but after 3.0 Ga, eclogitic diamonds became prevalent, indicating the onset of subduction and modern plate tectonics (Shirey & Richardson, 2011). Moreover, xenon isotope data indicate that volatile recycling was minimal before 3.0 Ga. Also, a reduction in crustal growth occurred around this time and is attributed to the accelerated destruction of differentiated continental crust. The significant increase in the volume of rocks preserved from the end of the Archean, coinciding with other major changes at that time, likely resulted from subduction and continental collision and could be considered indicative of the period when plate tectonics became the prevailing tectonic regime on Earth and the onset of the Wilson cycle (opening and closing of oceanic basins) (Hawkesworth et al., 2020; Shirey & Richardson, 2011). These and related details are discussed extensively elsewhere in this book.

Plate tectonics has influenced nutrient and volatile cycling significantly throughout Earth history. It plays a crucial role in the carbon cycle by facilitating through subduction the return of atmospheric $CO_2$ sequestered in buried organic matter and carbonate rocks. The carbonate-silicate cycle functions as a long-term thermostat, regulating surface temperatures by enhancing chemical reactions and in particular temperature-dependent weathering of silicate rocks, which increase $CO_2$ drawdown (Berner et al., 1983; Walker et al., 1981; Windley et al., 2021). Similarly, plate tectonics is essential for the $CH_4$ cycle, enabling the subduction and recycling of methane-rich sediments and driving volcanic outgassing. The different tectonic activity in Earth's early history could have impacted $CH_4$ cycling, thereby altering atmospheric composition in ways distinct from those operating today (Gerya, 2014; Mörner & Etiope, 2002). The presence of methane-rich fluid inclusions in olivine-bearing rocks confirms the generation of abiotic $CH_4$ reservoirs on Earth. This abiotic pathway for $CH_4$ production may have been even more critical in the early history of Earth because of the abundant mafic/ultramafic crust, which provided the necessary conditions for such chemical processes, prior to the emergence of methanogenic archaea (Etiope & Whiticar, 2019; Klein et al., 2019).

Both biotic and abiotic $CH_4$ production requires a reductant, typically $H_2$. One major abiotic source of $H_2$ in geological systems is serpentinization, whereby ultramafic rocks such as peridotite, rich in olivine minerals, react with water, leading to the formation of serpentine minerals, magnetite, and $H_2$-rich fluids. $H_2$ derived from serpentinization can be used for $CH_4$ production both abiotically and through the specific microbial pathway of hydrogenotrophic methanogenesis (Etiope, 2017; Schrenk et al., 2013; Zabranska & Pokorna, 2018). Abiotically, it reduces aqueous $CO_2$ to $CH_4$ via the Sabatier reaction, which typically occurs under high temperatures and pressures near hydrothermal vent systems on the ocean floor. However, most serpentinizing systems operate at lower temperatures, where this reaction is less common. At these lower temperatures, the Sabatier reaction may still proceed slowly, suggesting that under specific conditions, such systems could contribute to $CH_4$ production, albeit at a reduced rate compared to their high-temperature counterparts. Additionally, molecular hydrogen ($H_2$), in combination with carbon-bearing compounds such as $CO_2$ from various sources, can participate in low-temperature Fischer-Tropsch Type reactions (FTT), leading to the production of alkanes and various other organic molecules, including $CH_4$. Recent studies have suggested that graphite (C), carbon monoxide (CO), formic acid (HCOOH), and formate ($HCOO^-$) could also be potential precursors for abiotic $CH_4$ production (Etiope & Whiticar, 2019; Steele et al., 2022). Furthermore, during serpentinization, iron is oxidized, transforming from ferrous ($Fe^{2+}$) to ferric ($Fe^{3+}$) states, while the renewal of crust through plate tectonics continuously replenishes $Fe^{2+}$. This oxidation creates reducing conditions in the environment, facilitating the production of hydrogen gas, which in turn supports the production of reduced gas species such as $CH_4$. Thus, plate tectonics could also have played a crucial role in facilitating these processes by replenishing new reactants and by creating fractures and enhancing water-rock interactions over time, particularly at mid-ocean ridges and subduction zones where geothermal and volcanic $CH_4$ emissions can be prevalent (Guzmán-Marmolejo et al., 2013; Judd, 2003).

The volcanic $CH_4$ flux during the Archean represents another significant abiotic source. Magma production rates may have been up to 25 times greater than those observed on contemporary Earth, which would have substantially elevated volcanic $CH_4$ emissions and atmospheric $CH_4$ levels (Wogan et al., 2020). Nevertheless, as mentioned earlier, some researchers argue that abiotic $CH_4$ production would have been 'sluggish' and relatively inefficient due to the kinetic barriers involved in the reduction of $CO_2$ to methane. In particular, abiotic $CH_4$ production in hydrothermal environments often occurs slowly and requires specific conditions, such as the presence of catalysts like the minerals awaruite or chromite, to proceed more efficiently. As a result, these processes are considered less significant compared to the more efficient biological pathways for methane production (Bradley, 2016). Therefore, achieving substantial atmospheric $CH_4$ at levels that would have significantly impacted global climate and environmental conditions may have required biological additions by methanogenesis rather than purely abiotic production (Krissansen-Totton et al., 2018b).

Weathering of Archean rocks would also directly impact the conditions for biological $CH_4$ cycling. As rocks underwent chemical transformations, significant changes in the chemical composition of Archean oceans and the bioavailability of essential nutrients paved the way for the diversification of early microbial pathways (Lyons, Tino et al., 2024, and discussions below). For instance, weathering would release iron, dissolved phosphate from weathered apatite, elevated levels of chromium and manganese from the dissolution of olivine, along with other trace minerals that are vital for microbial metabolism (Hao et al., 2017). The redox-dependent relationship between phosphorus (P) and iron (Fe) is a key example. Throughout the Archean, iron oxide precipitates may have actively scavenged dissolved P, potentially influencing the availability of this critical nutrient for early microbial life and biogeochemical cycles (Bjerrum & Canfield, 2002; Herschy, etal., 2018; Konhauser et al., 2007; Rego et al., 2023). The availability of nutrients established a foundation for the development and proliferation of microbial life in anaerobic settings at that time.

- **The Role of Biology**

The Archean Eon witnessed the advent of microbial processes that played significant roles in shaping the planet's carbon cycling mechanisms. Carbon metabolisms serve as a crucial link between the organic and inorganic phases of the carbon cycle (Lyons, Tino et al., 2024, Reinhardt et al., 2024). The earliest metabolic pathways were anaerobic, employing a range of electron acceptors to respire organic carbon or degrade it via fermentation. Despite being less energy efficient than aerobic respiration, anaerobic processes underpin a substantial portion of the microbial biosphere today and in the past, in environments where $O_2$ is/was limited or absent. Of particular importance is methanogenesis, an energy metabolism in which carbon itself acts as the electron acceptor, resulting in the production of $CH_4$ (Lyons, Tino et al., 2024).

Many key metabolic pathways and enzymes, including those playing vital roles in $CH_4$ cycling (Fig. 2), originated billions of years ago and coevolved with environmental changes. These microbial activities were capable of influencing environments at various scales, from local ecosystems to the planetary atmosphere, leaving behind unique geochemical signatures and genomic evidence, which can provide valuable insights into Earth's early history (Lyons et al., 2015, Moore et al., 2018). Nevertheless, the ambiguity in isotopic signatures, along with challenges in preserving diagnostic organic molecules and the absence of fossil evidence for microbial communities over extended periods, impedes a comprehensive understanding of ancient microbial activities and their ecological implications and demands additional data types. This gap underscores the need for innovative methodologies, such as genomic studies of ancient organisms—including those involved in $CH_4$ cycling or ancestral lineages of $O_2$ producers—to deepen our understanding of their ecological impacts (e.g., Fournier et al., 2021; Wolfe & Fournier, 2018). Additionally, this emerging synthesis provides invaluable insights as we search for life beyond Earth.

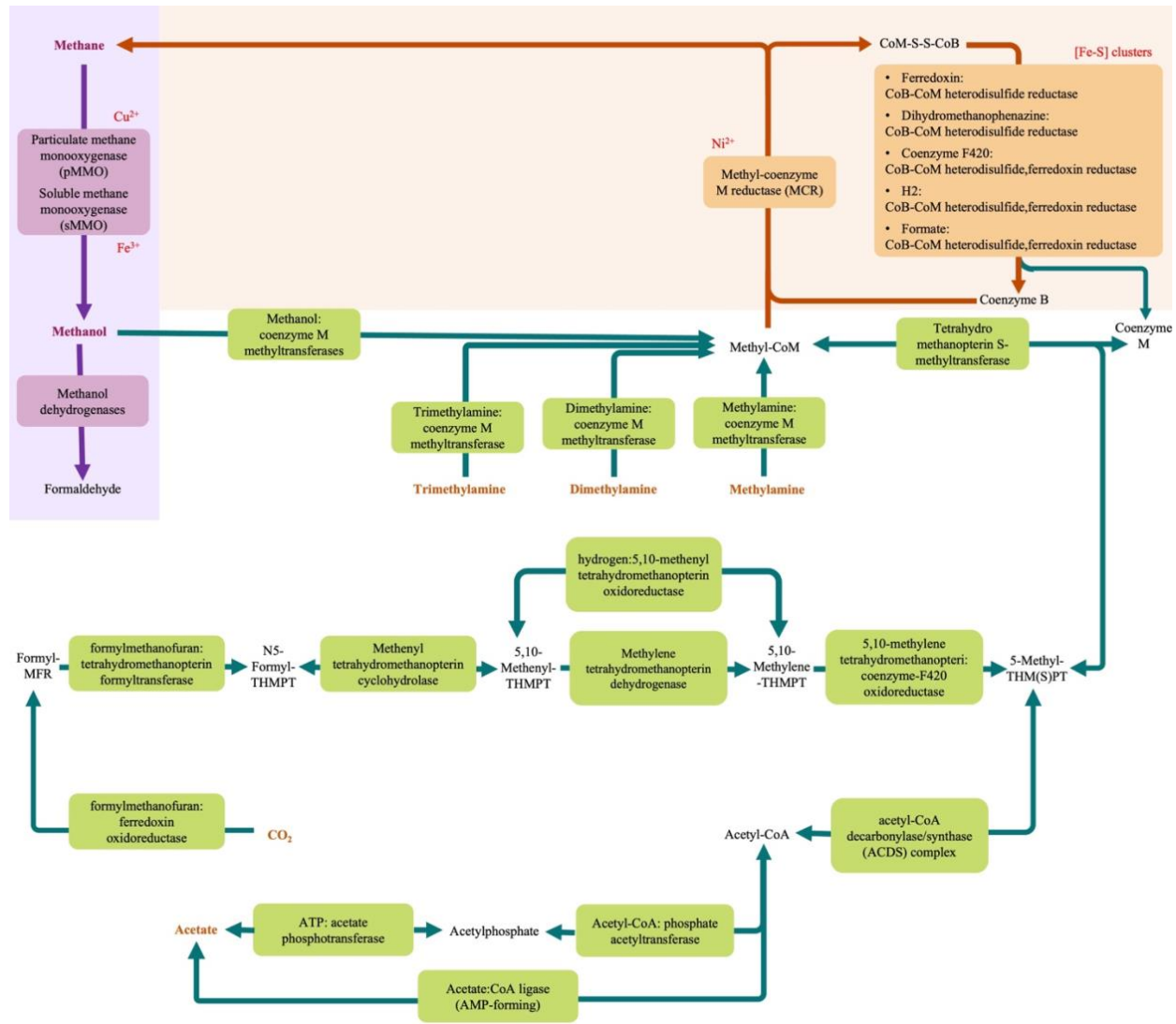

**Fig. 2: Methane metabolism - Reference pathway sourced from Kyoto Encyclopedia of Genes and Genomes (KEGG).** This figure illustrates the enzymatic pathways, including main substrates, metallic cofactors, and intermediates, involved in methane metabolism. The orange section outlines methanogenesis with key enzymes responsible for $CH_4$ production, such as methyl-coenzyme M reductase (MCR), which is vital for the final step of $CH_4$ synthesis. This process is crucial for the anaerobic production of $CH_4$, occurring in microbial environments such as wetlands and the guts of ruminants today, and it is also thought to have been prevalent on early Earth, supporting various anaerobic ecosystems. The purple section identifies methanotrophy, listing key enzymes critical for the first step in the aerobic oxidation of methane, including Particulate Methane Monooxygenase (pMMO) and Soluble

Methane Monooxygenase (sMMO) that oxidize $CH_4$ to methanol ($CH_3OH$). This process is a significant sink for atmospheric $CH_4$ and facilitates microbial energy production in oxygen-rich environments.

- **Methanogenesis and Methanotrophy**

The early Earth, prior to the rise of atmospheric $O_2$, provided a favorable environment for anaerobes, including methanogens. Given the widespread anoxic conditions, microbial methanogenesis was likely one of the most pervasive metabolic pathways at this time. As such, methanogens likely played a key role in shaping the Archean climate by releasing significant levels of $CH_4$ through various methanogenic pathways (Catling & Zahnle, 2020; Garcia et al., 2022; Lepot, 2020). Later in Archean, methanotrophs played an important role by oxidizing $CH_4$, either aerobically in the presence of oxygen or anaerobically using other electron acceptors (Cui et al., 2014; Guerrero-Cruz et al., 2021). This balance between microbial methane production and consumption was critical in regulating atmospheric $CH_4$ levels during the Archean.

- **Microbial methanogenesis**

Biogenic $CH_4$ results either from the anaerobic decomposition of organic matter or as a direct metabolic product of methanogenic archaea. Methanogens are obligate producers of $CH_4$ and do not grow using alternative electron acceptors for respiration (Lyu et al., 2018; Stafford et al. 1980). Biogenic $CH_4$ fluxes during the Archean therefore depended upon the availability of reductant ($H_2$) and biomass, as well as nutrients to support microbial populations. Methanogens are a diverse group of Archaea that utilize various substrates for $CH_4$ production under different environmental conditions (Costa & Leigh, 2014; Karekar et al., 2022; Kurth et al., 2020). They are currently classified into eight well-established orders, each supported by at least one cultured species. These orders span three primary phyla: Euryarchaeota (which includes Methanococcales, Methanopyrales, and Methanobacteriales), Halobacterota (comprising Methanomicrobiales, Methanocellales, Methanonatronarchaeales, and Methanosarcinales), and Thermoplasmatota (home to Methanomassiliicoccales). Ongoing metagenomic studies have recently uncovered additional methanogenic taxa within novel phyla, thereby broadening our understanding of their ecological distribution and evolutionary diversity. Euryarchaeota include mesophilic, thermophilic, and hyperthermophilic species that use $H_2$, $HCOO^-$, and sometimes CO as electron donors. They employ the Wolfe Cycle for rapid growth and energy generation, with some capable of nitrogen fixation. Halobacterota has a broader substrate range, including $CO_2/H_2$, $CH_3COO^-$, and methylated compounds. Notable adaptations include cold tolerance and the ability to metabolize complex compounds like coal-derived methoxylates. Methanomassiliicoccales and newer discoveries are primarily methylotrophic, growing under mesophilic conditions and in high-salinity environments (Lyu et al., 2018) (Table 1).

Although different methanogens utilize various substrates, the three main pathways of microbial $CH_4$ production are hydrogenotrophic, acetoclastic, and methylotrophic methanogenesis (Kurth et al., 2020). The first pathway, hydrogenotrophic methanogenesis, reduces $CO_2$ to $CH_4$ utilizing $H_2$ as the electron donor. The acetoclastic pathway involves splitting acetate ($CH_3COO^-$) into $CH_4$ and $CO_2$. And finally the methylotrophic pathway performs the dismutation of methylated compounds such as methanol, methylamines, and methyl sulfides (Ferry & Kastead, 2007; Garcia et al., 2022). Each methanogenesis pathway consists of multiple steps and requires numerous genes encoding several enzymes, along with both specialized organic and metal cofactors.

Hydrogenotrophic methanogenesis involves a complex series of reactions where $CO_2$ is sequentially reduced to $CH_4$ through single-carbon intermediates, each requiring specific cofactors. This process is known as the aforementioned Wolfe cycle. In this pathway, the newly synthesized methyl group is transferred to coenzyme M (CoM). The required reducing power is provided by electrons sourced from $H_2$ or organic compounds like $HCOO^-$ (Costa & Leigh, 2014; Zabranska & Pokorna, 2018). In methylotrophic methanogenesis, the transfer of the methyl group from methylated compounds (such as methylamines) to CoM is facilitated by specific sets of enzymes and corrinoid cofactors, which contain a central cobalt ion. In acetoclastic methanogenesis, the methyl group is derived from $CH_3COO^-$ through an acetyl-CoA intermediate. This intermediate is subsequently cleaved by the carbon monoxide dehydrogenase complex, resulting in the formation of $CO_2$ and a methyl group that is transferred to CoM (Lyons, Tino et al., 2024; Paulo et al., 2020).

Regardless of enzymatic pathway and substrate, all methanogenesis converges on a highly conserved step for the final reduction of the methyl group to $CH_4$. Necessary for energy generation, this final step entails the reduction of a methyl group bound to coenzyme M ($CoM-S-CH_3$ or methyl-CoM), using coenzyme B (HS-CoB). This reduction results in the production of $CH_4$ and the formation of heterodisulfide (CoM-S-S-CoB). The enzymes $N^5$-methyl-$H_4$SPT M methyltransferase (MtrABCDEFGH or MTR) and methyl-coenzyme M reductase (McrABG or MCR) mediate this process (Borrel et al., 2019; Garcia et al., 2022) and are integral to the methanogenic biochemistry of archaea.

Despite the modest amount of free energy obtained through these reactions, methanogenesis provides sufficient energy for growth in environments where other metabolisms are infeasible (such as those lacking $O_2$ and sulfate, see discussions below; Alfano & Cavazza, 2020; Kurth et al, 2020; Thauer, 1998). Based on standard free energy changes ($\Delta G'^\circ$), the most energetically favorable reaction for $CH_4$ production is from $H_2$ and $CO_2$ (−135.6 kJ/mol), followed by formate, methanol, and methylated compounds such as methylamines and dimethylsulfide as substrates, with acetate being the least favorable (−31.0 kJ/mol) (Garcia et al., 2000). Pathways such as acetoclastic methanogenesis are notably constrained by the limited availability of specific organic compounds required for the process (Lyons, Tino et al., 2024). Consequently, the efficiency and prevalence of these methanogenic pathways are directly influenced by the environmental concentration of these substrates. Some methanogens form syntrophic relationships with fermentative bacteria, providing the required substrates and enabling the complete degradation of organic matter and $CH_4$ production. This mutual dependence enhances the overall metabolic efficiency and stability of the microbial community in such settings (Conrad, 2020; Sieber et al., 2012).

**Table 1: Classification and characteristics of known methanogens.** Resolving the phylogenetic boundaries between different groups of methanogens remains challenging. Several methanogen orders are underrepresented, with their characteristics largely inferred from a single genus. The discovery and isolation of additional strains will be essential for deepening our understanding of their diversity, adaptations, and for improving the robustness of their taxonomic classification (Further information can be found in Feehan et al., 2023; Kurth et al, 2020; Lyu et al., 2018; Lyu & Lu, 2018a; Lyu & Lu, 2018b; Mand & Metcalf, 2019; Wang et al., 2021; Xie et al., 2024).

| Phylum | Order | Metabolic Traits | Ecological Adaptations |
|---|---|---|---|
| **Euryarchaeota** | Methanococcales | Use $CO_2$ and $H_2$ or alternatively formate as an electron donor. | Prefer mesophilic to hyperthermophilic conditions and have been detected in hypersaline environments. |
| | Methanopyrales | Perform $CO_2$ reduction using $H_2$. | Hyperthermophilic (adapted to extreme heat with the ability to perform methanogenesis at temperatures exceeding 100 °C, likely due to a notably high GC content in the genome) and can grow in slightly acidic to neutral PH levels. |
| | Methanobacteriales | Utilize a diverse range of substrates for methanogenesis, including $H_2$ and $CO_2$, and in some cases, members of this group can reduce methanol and methylamine, and are able to use formate, CO, or secondary alcohols as electron donors. | Typically prefer mesophilic to thermophilic environments and in some cases hyperthermophilic conditions. They live in slightly acidic to alkaline PH levels. |
| **Halobacterota** | Methanomicrobiales | Reduce $CO_2$ with $H_2$, and in some cases utilize, formate and secondary alcohols as alternative electron donors. | Most species are mesophilic, though some are adapted to cold (psychrophilic) or hot (thermophilic) conditions. They generally grow near neutral PH; however, some exceptions can grow in acidic or alkaline environments. |
| | Methanocellales | Use $CO_2$ and $H_2$ or alternatively formate. | Optimally grow in mesophilic conditions and neutral PH but are capable of surviving in moderately acidic environments. |
| | Methanosarcinales | Utilize the broadest range of substrates for methanogenesis, such as $H_2$ and $CO_2$, acetate, and various methylated compounds including but not limited to methanol and methylamines. Interestingly, some are also capable of growth and methanogenesis using methoxylated aromatic compounds (MACs), such as methoxy-benzoate (methoxydotrophic methanogenesis). | Exhibit versatile environmental adaptations, ranging from cold-tolerant (psychrotolerant) to mesophilic to thermophilic. This group includes moderate to extreme halophiles and halotolerant species. Most of them prefer neutral pH, though some can inhibit moderately acidic or alkaline environments. |
| | Methanonatronarchaeales | Reduce methyled compounds using electrons derived from $H_2$. They can also use formate as an electron source. | Thrive in extremely saline environments, close to salt saturation. Most of them are thermophiles and alkaliphiles. |
| **Thermoplasmatota** | Methanomassiliicoccales | Reduce methylated compounds such as methanol with electrons derived from $H_2$. It is suggested that some members of this group may depend on methanogenesis using $H_2$ and $CO_2$. Some studies suggest that some of them can also use acetate or reduce tri-, di- and monomethylamine with $H_2$. | Inadequately characterized, but likely prefer mesophilic environments, and can survive in moderately acidic to alkaline conditions. |

How old is microbial methanogenesis, and what are the relative antiquities of its different pathways? Although the corrinoid-dependent and acetoclastic pathways appear to be simpler forms of microbial methanogenesis, they have a narrower taxonomic distribution, being found only in two orders: Methanomassiliicoccales and Methanosarcinales. This limited distribution suggests that these pathways may represent more recent evolutionary adaptations rather than ancestral forms of methanogenesis. In contrast, the widespread taxonomic distribution of the hydrogenotrophic pathway suggests that it is likely the ancestral form of methanogenesis (Adam et al., 2022; Berghuis et al., 2018; Borrel et al., 2013; Lyons, Tino et al., 2024). This pathway is found across nearly all orders of methanogens, with the exception of Methanomassiliicoccales, which primarily rely on methylotrophic methanogenesis (Borrel et al., 2014). Methylotrophic methanogenesis may have spread between Methanosarcinales and Methanomassiliicoccales through ancient horizontal gene transfer (HGT) events (Adam et al., 2022; Berghuis et al., 2018). A related evolutionary question is the history of the Wood-Ljungdahl Pathway (WLP) for carbon fixation. This presumably ancient pathway is used by some bacteria and many archaea, including methanogens. Specifically, hydrogenotrophic methanogenesis is associated with the methyl branch of the WLP, and the two pathways may have co-evolved (Borrel et al., 2016). The deep evolutionary history of methanogenesis and its relationship with other carbon-reducing metabolisms, such as acetogenesis, remain a topic of active debate within the scientific community, with ongoing research into the origins of these metabolic pathways (Homann et al., 2018; Martin & Russell, 2007; Mei et al., 2023; Sousa et al., 2013).

- **Phylogenomic and Geochemical Evidence for Ancient Methanogenesis**

Recent phylogenetic and molecular clock studies have advanced our understanding of the evolutionary timeline of major microbial lineages significantly. These studies provide a wide range of time estimates for the emergence of methanogens, spanning from as early as 4.49 Ga to as recent as 2.4 Ga or even later (Battistuzzi et al., 2004; Battistuzzi & Hedges, 2009; Marin et al, 2017; Sheridan et al., 2010). Given the inherent uncertainty in relaxed molecular clocks, and the wide variety of datasets, models, and calibrations used, results of molecular clock studies often vary substantially, and care must be taken in presenting and comparing their findings. With that caveat in mind, here we present the stated findings of a few recent attempts at dating the antiquity of methanogens.

One molecular clock analysis using a dataset of 29 highly conserved, primarily ribosomal, universally distributed proteins suggests that LUCA (the last universal common ancestor of cellular life) predated the end of the proposed Late Heavy Bombardment (>3.9 Ga). This approach placed the estimate for the age of crown Euryarchaeota, which includes methanogens, between 2.881 and 2.425 Ga (Betts et al., 2018). Another study, based on analyses of ribosomal sequences within a species tree framework, estimated the divergence of methanogenic lineages to have occurred ~3.53 ± 0.164 Ga (Wolfe & Fournier, 2018). Their molecular clock studies incorporating the HGT of proteins from Group I methanogens to Cyanobacteria —the Structural Maintenance of Chromosomes (SMC) and Segregation and Condensation protein families (ScpA/ScpB)— permitted the use of cyanobacterial fossil calibrations to date methanogen evolution. These analyses recover methanogenic Euryarchaeota diversifying no later than 3.51 Ga (Wolfe & Fournier, 2018). One major factor impacting the inference of the age of methanogenesis is where on the tree of life it originated. While arguments from parsimony generally place methanogenesis evolving within Euryarchaeota, a proposed alternative rooting of Archaea would place methanogenesis as an ancestral trait of all Archaea (Raymann et al., 2015). This hypothesis is further supported by the discovery of methane-cycling genes in Bathyarchaeota (a lineage within the TACK clade of Archaea), which may indicate the presence of methanogenesis outside the Euryarchaeota (Borrel et al., 2016).

A Paleoarchean age for methanogenesis is further supported by geochemical evidence, notably the discovery of isotopically light $CH_4$ (<-56‰) within fluid inclusions from the ~3.46 Ga Dresser Formation in Western Australia (Ueno et al., 2006). However, it is crucial to consider the potential for abiotic sources contributing to the observed $CH_4$ and its isotopic signature in fluid inclusions (Etiope & Lollar, 2013; Lollar & McCollom, 2006), necessitating a cautious interpretation of these findings. Potential evidence of methane cycling is also preserved in isotopically light Paleoarchean microfossils (Flannery et al., 2017; Schopf et al., 2017), though it remains uncertain whether they were associated with biotic or abiotic $CH_4$. All in all, current phylogenomic and geochemical evidence support methanogenesis dating back to the earliest diversification of the Euryarchaeota, likely before 3.6 Ga (from 3.5 to 3.9 Ga) (reviewed in Lyons, Tino et al., 2024; Wolfe & Fournier, 2018), and, as such, it is likely one of the most ancient metabolic processes on Earth.

- **Microbial Methanotrophy**

Methane is metabolized by diverse groups of methanotrophic microorganisms through both aerobic and anaerobic pathways. Aerobically, methanotrophs utilize $O_2$ as the terminal electron acceptor to oxidize $CH_4$ (Guerrero-Cruz et al., 2021). In environments where $O_2$ is scarce or absent, Anaerobic Oxidation of Methane (AOM) has likely played a critical role in controlling atmospheric $CH_4$ levels throughout geologic history. ANaerobic MEthanotrophs (ANME) utilize $CH_4$ as an electron donor and engage various sulfur compounds—such as sulfate ($SO_4^{2-}$), elemental sulfur ($S^0$), thiosulfate ($S_2O_3^{2-}$), or tetrathionate ($S_4O_6^{2-}$)—as electron acceptors in sulfate-dependent anaerobic methane oxidation (S-DAMO) (Cui et al., 2014; Cassarini et al., 2017; Gwak et al., 2022). Beyond sulfur compounds, nitrogen compounds like nitrite ($NO_2^-$) and nitrate ($NO_3^-$) serve crucial roles in anaerobic methane oxidation (N-DAMO) (Cui et al., 2014; Welte et al., 2016). Oxidized metals also act as important electron acceptors in this multifaceted process. Anaerobic methanotrophs have the capability to catalyze metal ion-dependent anaerobic methane oxidation (M-DAMO), effectively coupling AOM to metal reduction using $Fe^{3+}$ or $Mn^{4+}$ within methane-rich environments (Cui et al., 2014; Ettwig et al., 2016; He et al., 2017; Riedinger et al., 2014). This capability underlines the diverse mechanisms through which anaerobic methanotrophs contribute to $CH_4$ and carbon cycling and regulation in anoxic conditions, potentially influencing atmospheric $CH_4$ concentrations over extensive periods of geologic time. For instance, iron-dependent anaerobic oxidation of methane (Fe-AOM) may have significantly influenced the biogeochemical dynamics of the Archean oceans, a time characterized by prolific production and deposition of iron oxides often in massive iron formations formed within ferruginous (Fe-rich) oceans. This period was distinguished by a biosphere abundant in $CH_4$ yet markedly low in $SO_4^{2-}$ concentrations, conditions that likely enhanced the ecological and geochemical importance of Fe-AOM processes (Riedinger et al., 2014).

Anaerobic Methanotrophic archaea oxidize $CH_4$ through reverse methanogenesis using versions of the same enzymes present in methanogenic pathways (Glass & Orphan, 2012). A notable example is MCR, a key metalloenzyme present in both methanogens and specific groups of ANME. This enzyme is directly involved not only in the final step of $CH_4$ production but also in the initial step of $CH_4$ oxidation (Wang et al., 2021). Currently, all ANMEs are phylogenetically classified within the superphylum Euryarchaeota of the domain Archaea. They belong to the order Methanophagales, which consists of the ANME-1 cluster and the order Methanosarcinales, which includes the ANME-2 cluster (ANME-2a/b, ANME-2c, ANME-2d) and ANME-3 (Wang et al., 2022).

Aerobic methanotrophic bacteria, which can be classified into these major groups: Types I and X methanotrophs (Gammaproteobacteria), Type II methanotrophs (Alphaproteobacteria), and Type III methanotrophs (Verrucomicrobia) likely evolved later and began consuming $CH_4$ once oxygen was available in at least some environments. The Gammaproteobacteria class utilize the ribulose monophosphate (RuMP) pathway for formaldehyde assimilation, while the Alphaproteobacteria class rely on the serine pathway for formaldehyde assimilation. And Verrucomicrobia phylum can thrive in extreme environments, such as acidic or geothermal areas. Methylococcus capsulatus (Bath) (Type X) and Methylosinus trichosporium OB3b (Type II) are well-studied examples of these methanotrophs (Khider et al., 2021; Op den Camp et al., 2009). Aerobic methanotrophy in all aforementioned groups is facilitated by methane monooxygenases, which initially oxidize $CH_4$ to methanol. The ancestors of these organisms could have played a crucial role in reducing $CH_4$ concentrations in areas where $O_2$ started to accumulate, possibly shortly before and during the GOE (Kalyuzhnaya, 2019; Smith et al., 2011).

- **Phylogenomics and Geochemical Evidence for Ancient Methanotrophy**

The evolutionary history of methane oxidizing microbial metabolisms has not been studied as extensively as methanogenesis, although the genomic and geological record have provided some clues to their antiquity. One study suggests that the significant depletion in $^{13}C$ (-57‰) of ~2.7 Ga kerogen ($\delta_{ker}$) indicates a substantial increase in the availability of oxidized electron acceptors (such as $O_2$, $SO_4^{2-}$, and $NO_3^-$), driving increased rates of bacterial aerobic methanotrophy (Eigenbrode & Freeman, 2006). Anaerobic methanotrophy by consortia of archaea and bacteria is also possible and would similarly be indirectly enabled by the rise in $O_2$ (Eigenbrode et al., 2008). Recent comparative genomic analyses and molecular clock dating support the emergence of the major ANME lineages between 2.66 and 1.88 Ga, during the late Archean or early Proterozoic. This timing suggests a potential correlation with significant geobiological events, such as the GOE and the Huronian Glaciation, potentially contributing to the latter via a reduction of atmospheric $CH_4$ levels (e.g., Bekker, 2023; Wang et al., 2022).

- **Environmental Changes: The Bioavailability of Metallic Cofactors, Sulfate and Oxygen**

Fluctuations in the availability of trace metals, driven by environmental changes, likely played a pivotal role in the evolution and prevalence of microbial metabolisms, particularly those involved in $CH_4$ cycling. For instance, the Ni flux from geological sources has varied over time. A notable decline in ultramafic volcanism could have triggered a 'nickel famine,' directly suppressing the activity and biomass of methanogens, which rely on Ni for essential enzymatic processes. Nickel is crucial for several cofactors and enzymatic steps in methanogenesis, including the function of

F$_{430}$, a nickel-containing cofactor in MCR. Nickel is also essential for the functioning of hydrogenases, which participate in the initial stages of the hydrogenotrophic pathway by facilitating the reduction of $CO_2$ with hydrogen ([Scheller et al., 2010](#); [Thauer et al, 2010](#)). The decline in Ni availability could have constrained biogenic $CH_4$ production, potentially influencing the overall $CH_4$ budget and contributing to shifts in atmospheric composition and specifically a decreasing capacity for greenhouse warming that could have triggered widespread glaciation ([Bekker, 2023;](#) [Konhauser et al., 2015](#); [Lepot, 2020](#); [Moore et al., 2017](#)). However, some recent studies on the fluctuations in $O_2$ levels over time suggest that the interplay between atmospheric oxygen and the availability of essential metals like Ni may have been more complex and variable than previously thought. These findings indicate that the environmental controls on $CH_4$ production, including trace metal availability, were likely influenced by a broader range of factors, complicating our understanding of early Earth's atmospheric evolution ([Chen et al., 2022](#)).

There are also different types of iron-containing functional enzymes in microbial methanogenesis ([Ma et al., 2024](#)). Thus, the bioavailability of transition metals plays a crucial role in the microbial cycling of $CH_4$. Methanogenic archaea rely not only on Ni and Fe but also on other metallic cofactors such as cobalt (Co), and in some pathways, molybdenum (Mo), tungsten (W), and zinc (Zn) are involved. Various studies have highlighted the low bioavailability of Ni, Fe, and Co as significant limiting factors in microbial methanogenesis ([Glass & Orphan, 2012](#); [Ma et al., 2024](#)). Moreover, in aerobic oxidation of $CH_4$, the initial step relies on either Cu or Fe, with a strong dependence on the bioavailability of Cu. Subsequent steps involve the incorporation of additional Fe, Cu, and Mo ([Glass & Orphan, 2012](#)). Table 2 provides an overview of the metal cofactors involved in microbial methane production and consumption.

**Table 2: Metal Cofactors in Microbial Methane Production and Consumption.** Methanogenesis is one of the most metal-dependent enzymatic pathways in microbial metabolism. Metallic cofactors play key roles in both aerobic and anaerobic methanotrophy as well (**[Glass & Orphan, 2012](#); [Ross & Rosenzweig et al., 2016](#)**).

| Metal Cofactor | Enzyme/Protein | Function |
|---|---|---|
| Nickel (Ni) | Methyl-coenzyme M reductase (MCR) | Key Enzyme in methanogenesis, catalyzes the last step of $CH_4$ production and the first step of AOM via reverse methanogenesis. |
| | Carbon monoxide dehydrogenase/acetyl-CoA synthase complex (Cdh) | Cleaves the $CH_3$ group off acetyl-CoA and transfers it to $CH_3$-H$_4$SPT. |
| | Energy-conserving hydrogenase complexes (Ech/Eha/Ehb/Mbh) | Energy-converting hydrogenases that oxidize $H_2$ and reduce electron carriers. |
| | F420-reducing Hydrogenase (Frh) | Coenzyme F420-reducing enzyme. |
| | Ni–Fe hydrogenases (Vh(o/t)/Mvh) | Oxidize $H_2$ to provide electrons for reducing electron carriers |
| Cobalt (Co) | Methanol-coenzyme M methyltransferase (Mta) | Catalyzes the transfer of a methyl group from methanol to CoM to form methyl-CoM (a key intermediate in methylotrophic methanogenesis). |

| | CH$_3$-H$_4$M(S)PT-coenzyme M methyltransferase (Mtr) | Catalyzes the transfer of a methyl group from methyl-tetrahydromethanopterin (CH$_3$-H$_4$MPT) to CoM, forming methyl-CoM (a key intermediate in methanogenesis). |
|---|---|---|
| | Carbon monoxide dehydrogenase/acetyl-CoA synthase complex (Cdh) | Other than Ni, it also involves Co as part of its structure for catalyzing reactions in methanogenesis. |
| Tungsten (W) | W-containing formylmethanofuran dehydrogenase (Fwd) | Initiates the first step in methanogenesis by catalyzing the reduction of CO$_2$ in environments where W is more abundant and bioavailable. |
| Zinc (Zn) | Methanol-coenzyme M methyltransferase (Mta) | In addition to the Co cofactor, it contains a zinc atom, which is essential for its catalytic activity. |
| | Heterodisulfide reductase (Hdr) | Catalyzes the reduction of CoM-S-S-CoB to form HS-CoM and HS-CoB during the second-to-final step of methanogenesis, regenerating the cofactors required for CH$_4$ production. |
| Molybdenum (Mo)* | Nitrogenase | Nitrogen fixation |
| | Mo-containing formylmethanofuran dehydrogenase (Fmd) | Initiates the first step in methanogenesis by catalyzing the reduction of CO$_2$ in environments where Mo is more abundant and bioavailable. |
| Iron (Fe) | Several enzymes involved in microbial methanogenesis – containing [Fe-S] clusters | Key roles in different methanogenesis pathways. |
| | Soluble methane monooxygenase (sMMO) | Aerobic CH$_4$ oxidation |
| Copper (Cu) | Particulate methane monooxygenase (pMMO) | Aerobic CH$_4$ oxidation |

*Molybdenum is a versatile metal cofactor, crucial not only for nitrogen fixation in nitrogenase but also potentially involved in enzymes related to CH$_4$ metabolism under certain conditions. This can include indirect roles where nitrogen fixation influences microbial ecosystems that subsequently impact CH$_4$ dynamics.

Moreover, other geological factors would have influenced the biotic part of the CH$_4$ cycle, especially by the late Archean, as atmospheric oxygen levels began to increase at least locally. Aside from the availability of metal cofactors, a prime example is the role of sulfate (SO$_4^{2-}$). Specifically, very low Archean marine SO$_4^{2-}$ (Crowe et al., 2014; Habicht et al., 2002) would have favored the production of CH$_4$, as the more energetic microbial process of sulfate reduction would have otherwise outcompeted methanogens for reductants (Liu, 2012; Susanti & Mukhopadhyay, 2012). Moreover, sulfate-based anaerobic oxidation of CH$_4$ would be minimal when sulfate is scarce. The presence of O$_2$ also challenges CH$_4$ production and preservation by restricting the abundance of the necessary anaerobic environments; facilitating the aerobic oxidation of CH$_4$; and, as with sulfate, outcompeting methanogens for reductants. Sulfate and oxygen were vanishingly low during the Archean, and the deleterious consequences of concomitantly rising SO$_4^{2-}$ and O$_2$ became a factor later, during the Proterozoic (Olson et al., 2016). Thus, the controls governing both abiotic and biotic CH$_4$ flux to the atmosphere, as well as its preservation, have collectively limited abundant methane accumulation in the atmosphere since the onset of the GOE.

A comprehensive overview of the biogeochemical reactions involved in CH$_4$ cycling is provided in Table 3, including photochemical processes, hydrothermal reactions, and microbial activities. These reactions and their interactions elucidate the complex interplay between various molecules in different geological settings and diverse biological pathways that shaped the $CH_4$ cycle on early Earth.

Table 3: Abiotic and biotic methane cycling reactions on early Earth (Cui et al., 2014; Etiope & Lollar, 2013; Kurth et al., 2020; Houghton et al., 2019; Levine et al., 1982; Schwander et al., 2023; Zahnle et al., 2020)

| Category | Reaction* | Process Description |
|---|---|---|
| Methane Photochemistry | $CH_4 + hv \rightarrow CH_2 + H_2$<br>Or $CH_4 + hv \rightarrow CH_3 + H$ | Photolysis at extreme UV wavelengths ($\lambda \leq 145$ nm). High-energy photons break down $CH_4$. |
| | $H_2O + hv \rightarrow H + OH$<br>$CH_4 + OH \rightarrow CH_3 + H_2O$ | Water photolysis at longer UV wavelengths ($\lambda \leq 240$ nm). UV light splits water molecules. Methane reacts with hydroxyl radicals produced from water photolysis. |
| | $CH_3 + CH_3 + M \rightarrow C_2H_6 + M$ | In an $O_2$-depleted atmosphere, methyl radicals combine to form ethane with the help of a third body. |
| Abiotic Methane Production | Extraterrestrial $CH_4$ synthesis | Meteorites played a key role in facilitating this primordial process. |
| | $Fe + H_2O \rightarrow FeO + H_2$<br>$CO_2/CO + H_2 \rightarrow CH_4$ | Reaction of iron with water produces hydrogen, often from iron-rich impactors. Hydrogen produced from the iron-water reaction can react with carbon gases such as CO or $CO_2$ to form $CH_4$. |
| | $Fe^{2+}$ (in rock) + $H_2O$ (liquid) $\rightarrow Fe_3O_4$ (magnetite) + $H_2$ (in solution)<br>$CO_2/CO + H_2 \rightarrow CH_4$ | During serpentinization (near hydrothermal vents) ultramafic rocks (iron-bearing minerals such as olivine and/or pyroxene) react with water to produce $H_2$, which can then react with carbon gases such as CO or $CO_2$ to produce $CH_4$. |
| | $CO_2 + 4H_2 \rightarrow CH_4 + 2H_2O$<br><br>$CO_2 + H_2 \rightarrow CO + H_2O$<br>$CO + 3H_2 \rightarrow CH_4 + H_2O$ | Fischer-Tropsch Type (FTT) Reactions. For example, one-step methanation (Sabatier reaction), or two-step processes (reverse water-gas shift reactions) which can occur either after or independent of serpentinization), converting CO or $CO_2$ and $H_2$ to $CH_4$. |
| Biotic Methane Production | $CO_2 + 4H_2 \rightarrow CH_4 + 2H_2O$ | Hydrogenotrophic Methanogenesis. Microbial process using $CO_2$ and $H_2$ to produce $CH_4$. |
| | $CH_3COOH \rightarrow CH_4 + CO_2$ | Acetoclastic Methanogenesis. Microbial process using acetate to produce $CH_4$. |
| | $4CH_3OH \rightarrow 3CH_4 + CO_2 + 2H_2O$ | Methylotrophic Methanogenesis. Microbial process using methanol to produce $CH_4$. |
| Methane Consumption | $CH_4 + O_2 + 2e^- + 2H^+ \rightarrow CH_3OH + H_2O$ | Oxidation by $O_2$. An enzymatic process in which methane reacts with oxygen to produce methanol as the first step in methanotrophy in aerobic environments. |
| | $CH_4 + SO_4^{2-} \rightarrow HCO_3^- + HS^- + H_2O$ | Oxidation by sulfate ($SO_4^{2-}$). Anaerobic process where $CH_4$ is oxidized using $SO_4^{2-}$. |
| | $CH_4 + 8Fe(OH)_3 + 15H^+ \rightarrow HCO_3^- + 8Fe^{2+} + 21H_2O$ | Oxidation by iron oxides. Methane is oxidized by Fe(III). |

| | |
|---|---|
| $CH_4 + 4MnO_2 + 7H^+ \rightarrow HCO_3^- + 4Mn^{2+} + 5H_2O$ | Oxidation by manganese oxides. Methane is oxidized by Mn(IV). |
| $5CH_4 + 8NO_3^- + 8H^+ \rightarrow 5CO_2 + 4N_2 + 14H_2O$ | Oxidation by nitrate ($NO_3^-$). Methane is oxidized using nitrate, releasing nitrogen gas. |
| $3CH_4 + 8NO_2^- + 8H^+ \rightarrow 3CO_2 + 4N_2 + 10H_2O$ | Oxidation by nitrite ($NO_2^-$). Methane is oxidized using nitrite, releasing nitrogen gas. |

*Some reactions are Simplified.

## From an Anoxic to an Oxic Planet

Oxygenic photosynthesis stands as the most pivotal evolutionary innovation on our planet. It is believed to have originated from an ancestor of modern oxygen-producing cyanobacteria, likely an anoxygenic phototroph, and has profoundly shaped Earth's environmental evolution and life through time (Hohmann-Marriott & Blankenship, 2011; Shih, 2015; Ward et al., 2016). From microfossils and stromatolites to geochemical evidence, genomic insights, and molecular clock analyses, multiple lines of evidence support the presence of cyanobacteria by the late Archean and highlight the significance the evolution of oxygenic photosynthesis had on the early Earth (Fournier et al., 2021; Lepot, 2020; Sánchez-Baracaldo, 2022; Schirrmeister et al., 2016). With the onset of oxygenic photosynthesis and aerobic respiration, major biogeochemical transitions occurred, altering the planet's redox conditions. Even before $O_2$ levels increased in the atmosphere during the GOE, early aerobes likely exploited early $O_2$ accumulations within localized niches, particularly in the surface waters of productive regions (Olson et al., 2013). Growing evidence now suggests that free oxygen may have appeared in the surface ocean much earlier than the well-documented atmospheric oxygenation during the Proterozoic. Various geochemical proxies, including nitrogen isotopes (Garvin et al., 2009; Godfrey & Falkowski, 2009), the isotopic patterns of redox-sensitive elements (e.g., Kendall et al., 2010; Kendall, 2021), locally enhanced manganese (oxyhydr)oxide precipitation (Planavsky et al., 2010), and the isotopic composition and concentration patterns of molybdenum as a key proxy for studying past redox conditions (Anbar et al. 2007; Duan et al., 2010; Johnson et al., 2021; Scott et al., 2008; Scott et al., 2011) all argue for pre-GOE emergence of $O_2$ in Archean surface waters, and transiently in the atmosphere (as discussed in detail by Ostrander et al. in this volume).

The initial emergence of free oxygen in the atmosphere is particularly supported by the loss of mass-independent fractionation of sulfur isotopes (MIF-S) (Farquhar et al., 2000; Olson et al., 2013, Ossa Ossa et al., 2019). The production via photochemistry and preservation of these isotopic anomalies are favored only when atmospheric $O_2$ concentrations are vanishingly low. This process required significant sulfur gas emissions, comparable to modern volcanic sources, and abundant $CH_4$ or other reduced gases. MIF-S anomalies typify the Archean and provide an indirect argument for elevated atmospheric methane (Zahnle et al., 2006), which may have also facilitated the formation of organic hazes (Arney et al., 2016; Zerkle et al., 2012). Such records indicate low atmospheric $O_2$ levels and minimal oxidative weathering before the GOE. However, there were exceptions with sporadic, small amounts of oxygen, often referred to as "whiffs" of $O_2$ or Archean Oxidation Events (AOEs) (Anbar et al. 2007; Anbar et al., 2023; Kendall, 2021; Ostrander et al., 2022), during which small quantities of $O_2$ were temporarily introduced into Earth's atmosphere and oceans.

Although estimates indicate that Archean $O_2$ levels, at least in the atmosphere, were extremely low —likely less than $10^{-6}$ times present-day levels—, they appear to have been significantly higher than those expected from abiotic processes (Catling & Zahnle, 2020; Johnson et al., 2021). Molybdenum isotope data from the Sinqeni Formation in the Pongola Supergroup, South Africa, reveal an early presence of manganese oxides in Archean environments, which requires substantial free oxygen concentrations for formation. Although manganese oxides are rarely preserved in ancient rocks, the significant molybdenum isotope fractionation associated with Mo absorption onto these Mn oxides can provide evidence of early $O_2$ production dating back at least 2.95 Ga (Planavsky et al., 2014). However, more recent findings suggest that manganese oxidation could also occur under anaerobic conditions through abiotic and light-driven microbial processes (Daye et al., 2019; see Robbins et al., 2023, review), which challenges the strict requirement of $O_2$. Despite the possibility of alternative pathways, manganese enrichments and related geochemical signals remain a reliable proxy for Earth's oxygenation and suggest that oxygenic photosynthesis may have emerged hundreds of millions of years before the onset of permanent atmospheric oxygenation (reviewed in Robbins et al., 2023). Another line of evidence, analysis of redox-sensitive elements within the Fig Tree Formation from South Africa, implies significant $O_2$ levels in the shallow oceans around 3.2 Ga (Satkoski et al., 2015), which might be attributable to the microbial primary production at that time. These findings, along with numerous other pieces of evidence, such as potentially biogenic structures like stromatolites, microbial mats, and microfossils found in Archean and even earlier samples (Barlow et al., 2024; Chatterjee, 2023; Djokic et al., 2017; Dodd et al., 2017; Schopf, 1993; Schopf, 2006), support the early emergence of life on Earth; however, the earliest geochemical evidence of photosynthesizers has been a topic of significant debate over the past decades (Nutman & Friend, 2009; Nutman et al., 2016; Nutman et al., 2021; Zawaski et al., 2020; Zawaski et al., 2021).

Genomic studies provide an independent source of evidence and suggest that oxygenic photosynthesis could have evolved in early microbial communities several hundred million years prior to the GOE (Fournier et al., 2021; Sánchez-Baracaldo & Cardona, 2019). These results agree with some of the most reliable geochemical evidence for AOE, indicating that there may have been periods when early Earth had sufficient $O_2$ to support aerobic microbial life, particularly in the surface oceans. These events likely created oxygen oases that could have served as ecological niches for early oxygen-dependent organisms, setting the stage for a more permanent rise in atmospheric oxygen. Further, they indicate that the mechanisms for $O_2$ production were already in place and that the Earth system was gradually adapting to the presence of $O_2$ long before the GOE. Consequently, the GOE can be seen as the culmination of a broader First Redox Revolution (FRR) in Earth history, characterized by two or more earlier AOEs, which likely played a crucial role as precursors to the GOE (Olson et al., 2013; Ostrander et al., 2021; Taverne et al., 2020). Over time, these small-scale oxidation events during the Neoarchean (2.8-2.5 Ga) contributed to the conditions that eventually led to a more sustained and significant alteration in the oxidation state of the atmosphere and shallow oceans in Paleoproterozoic, around 2.4-2.3 Ga (Lyons et al., 2014; Lyons et al., 2021; Ossa Ossa et al., 2019) The rise in atmospheric oxygen levels likely occurred due to complex microbial interactions and feedback mechanisms following the rapid diversification of ancestral photosynthetic cyanobacteria, which could have taken place around the time of the GOE (Fournier et al., 2021). The metabolic activity of these early photosynthesizers transformed Earth's atmosphere into an oxic state, and the persistent increase and ultimately accumulation of $O_2$ in Earth's atmosphere introduced a fundamental paradox in life and evolution, bringing both challenges —particularly for anaerobes, including methane producers— and opportunities for the emergence of more complex life forms (Lyons, Tino et al., 2024; Reinhard & Planavsky, 2020; Taverne et al., 2020). For a more in-depth exploration of the fascinating history of oxygen on Earth, see the review by Ostrander et al. in this volume.

- **The Methane-Oxygen Interplay**

The transition to an oxidized planet radically and permanently changed the biogeochemistry of $CH_4$ as part of the carbon cycle. The evidence for such a change reaches back into the late Archean, with ~2.7 Ga BIFs showing a decline in their nickel to iron ratio. This decline could be attributed to a reduced flux of Ni to the oceans (as discussed above), due to cooling upper-mantle temperatures and decreased eruptions of nickel-rich ultramafic rocks. Thus, the decreasing availability of volcanic Ni, an essential metallic cofactor for methanogenic enzymes, connects the evolution of the mantle to changes in the atmospheric redox state (Konhauser et al., 2009). Additionally, if primary productivity and oxygenic photosynthesis significantly increased in the ocean, they could suppress anaerobic microbial activity by reducing the availability of electron donors such as $H_2$ and CO, likely leading to a decrease in atmospheric $CH_4$ levels. This process is hypothesized to operate via the production of OH radicals from microbially produced $O_2$, which act as a sink for biogenic $CH_4$ and other electron donors (Watanabe et al., 2023).

The onset of increased $O_2$ levels in the atmosphere does appear to correlate with a decrease in Earth's temperature and major periods of Paleoproterozoic glaciation (Huronian glaciations). These were the earliest known series of extensive climatic cooling events that took place 2.45-2.22 Ga during the Paleoproterozoic era and may have been the greatest glaciation in Earth history (Bekker, 2023; Hoffman et al., 1988). The timing of this cooling directly implicates $O_2$, supporting its role in reducing the methane greenhouse effect that is thought to have kept early Earth warm during the Archean (Kasting & Catling, 2003; Lyons, Tino et al., 2024). Despite this reduction, some studies suggest that $CH_4$ concentrations could have remained relatively high throughout much of the Proterozoic due to low concentrations of dissolved oxygen and sulfate in the deep oceans, and the recycling of organic matter through processes such as methanogenesis (Kasting, 2005). Another hypothesis posits that $CH_4$ concentrations during the Proterozoic were lower than previously estimated. This alternative perspective attributes the lower $CH_4$ levels primarily to a diminished flux from oceanic environments during the Proterozoic, due to higher rates of ANME and more competition for reductants by sulfate reducers (Lyons et al., 2021; Olson et al., 2016), along with photochemical loss under higher but still low atmospheric $O_2$. An emerging topic in studies of the GOE is the evidence for oscillatory redox conditions, as highlighted in recent research (e.g., Bekker et al., 2020; Poulton et al., 2021), which suggests that the atmospheric oxygen levels fluctuated above and below critical thresholds around the GOE. Such fluctuations in atmospheric oxygen levels often correlate with changes in greenhouse gas concentrations, particularly due to the oxidation of atmospheric $CH_4$, which may have contributed to the onset and retreat of glaciations. As discussed earlier, one interesting aspect of methane and oxygen interactions is that while $O_2$ can destroy $CH_4$, ozone's UV shielding from increased $O_2$ could have reduced methane's photochemical destruction (Lyons, Tino et al., 2024; Olson et al., 2016). These feedbacks between atmospheric oxygen, $CH_4$, and glaciations illustrate the delicate balance of early Earth's atmospheric system. A deeper understanding of these dynamics provides key insights into the long-term stability and variability of Earth's climate throughout geological time.

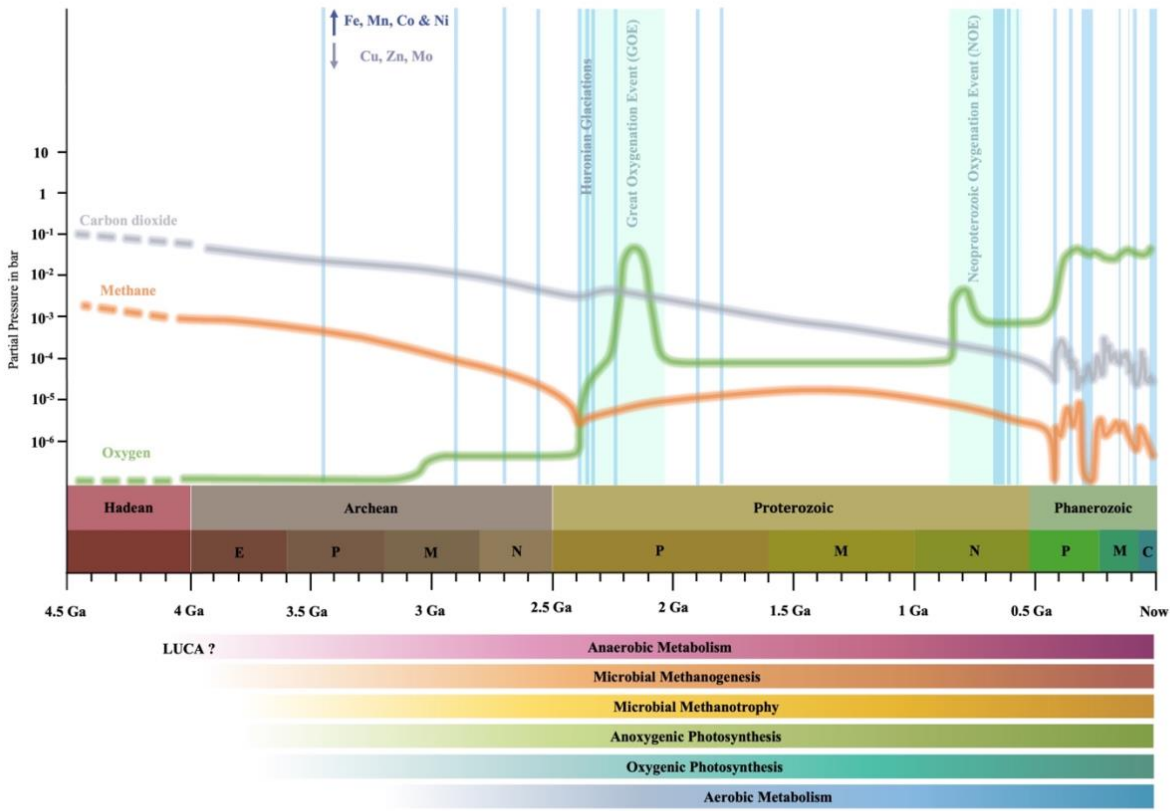

**Fig3. Evolving Earth and life through time:** Key elements of this co-evolution include shifts in atmospheric and oceanic composition, the emergence of various metabolic pathways, and significant geological and biological events. Carbon dioxide, methane, and oxygen have been key components of Earth's atmosphere throughout geological time. The gradual decline in $CO_2$ reflects a feedback process in the carbon cycle, driven by increasing solar luminosity. As oxygen levels increased, methane was initially oxidized but may have later been shielded by an ozone layer. Key events illustrated include major oxygenation events (GOE and NOE, highlighted in light green), periods of glaciation (blue lines marking intervals with glacial deposits that indicate episodes of glaciation), and the evolutionary development of diverse life forms over geologic time. Key microbial metabolic pathways—such as the onset of aerobic and anaerobic metabolism, including methanogenesis, methanotrophy, and anoxygenic and oxygenic photosynthesis—are depicted as vertical colored bars, illustrating their likely timing within this planetary context (Anbar, 2008; Catling & Zahnle, 2020; Fournier et al., 2021; Lyons, Tino et al., 2024).

## Challenges and Future Directions in Research

The Archean Earth was a planet very different from the one we know today. The atmosphere was anoxic and likely rich in greenhouse gases, while the hydrosphere was predominantly composed of warm, acidic and iron-rich oceans. However, against this backdrop, there was likely substantial environmental variability, with numerous transient events, such as impacts and periods of increased or decreased volcanism. The atmosphere and hydrosphere exhibited markedly low levels of $O_2$ and $SO_4^{2-}$, creating a reducing environment where methanogens had less competition from other metabolisms, such as sulfate reduction. In the absence of significant oxidants, $CH_4$, once produced, was likely to accumulate rather than be rapidly oxidized and converted back to $CO_2$.

Later, following the advent of oxygenic photosynthesis, subsequent increases in the concentrations of $O_2$ and $SO_4^{2-}$ would have significantly limited both the production and preservation of $CH_4$. More geochemical and genomic evidence constraining the extent of biological $CH_4$ production and consumption has the potential to substantially improve our understanding of Archean Earth and the transition to a more oxygenated planet. Further study of preserved Archean $CH_4$ using clumped isotope analysis ([Giunta et al., 2019](Giunta et al., 2019)) can distinguish between abiotic and biotic $CH_4$ sources, offering key geochemical evidence for the relative extent of $CH_4$ generation and consumption pathways. Additionally, studying trace metal redox proxies, such as molybdenum and iron, could significantly refine our understanding of the redox conditions in the Archean atmosphere and hydrosphere by providing indirect evidence of $O_2$ levels and oxidative events, which are critical for constraining early $CH_4$ cycling. These advanced techniques would offer an even deeper perspective on Archean $CH_4$ and related biogeochemical processes.

Although sedimentary records and genomic insights have been instrumental in deciphering the enigmatic Archean eon, offering a window into ancient climates, microbial evolution, and the narrative of biogeochemical cycles across geological and biological realms, it is important to acknowledge that our current perspectives are part of an ongoing dialogue within the scientific community, where established views are continually challenged and refined. The interplay of new analytical techniques and interdisciplinary research will likely lead to paradigm shifts, offering fresh insights into the Archean Eon and beyond. Comprehensive and parallel biogeochemical, genomic, and geological analyses, conducted by independent and collaborating research groups, will continue to be important in constructing this narrative. In particular, the biogenicity of any proposed ancient bio-structures or geochemical signals, and the phylogenetic methods for reconstructing the microbial natural history must be continually and rigorously assessed. To this end, future research directions may involve refining our understanding of specific geological processes that favor life, delving deeper into the intricacies of microbial interactions, and utilizing advanced technologies to extract more detailed information from ancient sedimentary and genomic records.

In closing, we hope that ongoing inquiries and future collaborations will further advance the boundaries of our understanding. The significance of interdisciplinary collaborations resonates throughout this narrative, emphasizing that a holistic understanding of Archean biogeochemical cycling requires a combination of knowledge from various disciplines. Moreover, the co-evolutionary relationship between Earth's geologic processes and biological systems during the Archean, encapsulated in the concept of the 'Archean tectonic-atmospheric-biotic cycle,' illustrates the deeply interconnected nature of planetary processes. In the realm of astrobiology and exoplanetary studies, the lessons learned from Earth's ancient biogeochemistry and methane dynamics offer valuable insights into potential abiotic and biotic mechanisms and controls on $CH_4$ production and consumption on distant worlds ([Stüeken et al., 2024](Stüeken et al., 2024)). As depicted, over Earth history, $CH_4$ transitioned from being a trace gas to an important component of the atmosphere, and then back to a trace gas, while $O_2$, which is now prevalent, was initially absent and subsequently scarce for a protracted period of Earth history. We caution, however, that while $CH_4$ was undoubtedly an important part of the Archean heat balance on Earth, we still know very little about how high and persistent atmospheric methane accumulations were during that time.

The intricate interrelationship between CH$_4$ and O$_2$, as discussed, is particularly fascinating and holds significant potential as a biosignature ([Schwieterman et al., 2018](); [Thompson et al., 2022]()), offering valuable insights into atmospheric and possible biochemical processes on other planets. While the simultaneous presence of O$_2$ and CH$_4$ in a planet's atmosphere is commonly viewed as a strong indicator of biological activity, it is also important to consider that this biosignature may have been challenging to detect through most of Earth's history (Reinhard et al., 2017). Similarly, low concentration intervals of both gases offered challenges, with notable exceptions being Archean biological CH$_4$ and Phanerozoic O$_2$. Importantly, though, Earth's likely persistence of undetectable atmospheric O$_2$ for billions of years following its biological origins is a classic example of a false negative in biosignature research (Reinhard et al., 2017). By broadening our understanding of such complex interplays, we enhance our ability to identify and understand the conditions necessary for life elsewhere within and beyond our solar system. As research progresses, the principle that "the past is the key to the future" becomes increasingly relevant. By uncovering new layers of Earth's early biogeochemical history, we deepen our understanding of the planet's past while simultaneously enhancing our knowledge of other potentially habitable worlds.